\documentclass[conference]{IEEEtran}
\IEEEoverridecommandlockouts
\usepackage{cite}
\usepackage{amsmath,amssymb,amsfonts}
\usepackage{algorithmic}
\usepackage{graphicx}
\usepackage{textcomp}
\usepackage{xcolor}
\usepackage{hyperref}
\def\BibTeX{{\rm B\kern-.05em{\sc i\kern-.025em b}\kern-.08em
    T\kern-.1667em\lower.7ex\hbox{E}\kern-.125emX}}
\begin{document}

\title{Agile Climate-Sensor Design and Calibration Algorithms Using Machine Learning: Experiments From Cape Point\\

\thanks{This work has been supported with funding from Sentech Soc Ltd, South Africa.}

}

\author{\IEEEauthorblockN{Travis Barrett}
\IEEEauthorblockA{\textit{Department of Electrical Engineering} \\
\textit{University of Cape Town}\\
Cape Town, South Africa \\
BRRTRA005@myuct.ac.za}
\and
\IEEEauthorblockN{ Amit Kumar Mishra}
\IEEEauthorblockA{\textit{Department of Electrical Engineering} \\
\textit{University of Cape Town}\\
Cape Town, South Africa \\
akmishra@ieee.org}

}

\maketitle

\begin{abstract}
In this paper, we describe the design of an inexpensive and agile climate sensor system which can be repurposed easily to measure various pollutants. We also propose the use of machine learning regression methods to calibrate CO2 data from this cost-effective sensing platform to a reference sensor at the South African Weather Service’s Cape Point measurement facility. We show the performance of these methods and found that Random Forest Regression was the best in this scenario. This shows that these machine learning methods can be used to improve the performance of cost-effective sensor platforms and possibly extend the time between manual calibration of sensor networks. 

\end{abstract}
\begin{IEEEkeywords}
Machine Learning, Sensor Calibration, Environment Monitoring, Random Forest, SVR
\end{IEEEkeywords}

\section{Introduction}
In the age of the Internet of Things (IoT), many projects are being created \cite{lee2015internet} to meet the ever-expanding web of systems that are all interconnected. With the goal of many of these projects to be as cost-effective as possible, low-cost sensors are often used. These sensors have varying levels of accuracy and performance characteristics that are influenced by the environment. Due to this, the performance of these sensors can vary heavily on their location which can produce erroneous data. Without a thorough investigation and methodology around sensor calibration, usually the data collected from the sensors is not reliable. For example, Bittner et al  \cite{bittner2022performance} discussed how they got their sensors calibrated to a high standard and then those were installed in Malawi. However, they observed how quickly the quality of the data decreased from the sensor network and eventually became mostly unusable.  This is a major pain-point in the current generation of ubiquitous sensing. It has been well described by Motlagh et al~\cite{Motlagh2020MassiveScaleAirQuality} that a high spatial density is required to effectively monitor the air quality of an area. Having a cost-effective solution that can be tailored to each location would allow for wide-scale monitoring to be more accessible. The system presented in this paper aims to meet these goals by offering a platform and methods to improve the accuracy of the sensors in the system. Motlagh et al~\cite{Motlagh2020MassiveScaleAirQuality} also state that large scale air quality monitoring would require opportunistic calibration transfer as the scale of deployments renders manual calibration infeasible. We aim to show the viability of machine learning to calibrate cost-effective sensor nodes to allow for large scale air quality monitoring. 

Calibration can be a complicated process. Laboratory calibration is a consistent method to properly calibrate sensors, however these laboratory conditions are not fully representative of the wide array of conditions the sensors may face \cite{rai2017end}. This will cause variance from the calibrated, expected, data curve. In order to re-calibrate the sensors as they will inevitably drift over time, the sensor will need to be returned to the laboratory to be calibrated again. The alternative to this is to take reference calibration equipment to the sensors in the field. This is a difficult and expensive method to calibrate the sensors in place as it will require a dedicated team to calibrate a sensor network. This process will not scale up to a large and/or remote sensor networks. In order to calibrate these sensors a reference sensor is required. These sensors are typically very expensive where accuracy of measurement is paramount. For instance, the Picarro G2401 sensor system used by the South African Weather Service in their Cape Point climate measurement station is costly and is not easily moved around. While it may not be possible to remove the need for reference equipment entirely, the greater the time between manual calibration of sensors will greatly reduce the cost and staff requirements for a given size of sensor network. The longer the sensors can have their data accurately calibrated through software methods, the lower the maintenance requirements of the network. 

As shown in Postolache~\cite{postolache2009smart} the ability for low cost sensors networks to be created has been around for many years. While this work shows the promise it was held back by the technology of the time. Making a system today gives the ability to use improved sensor technologies and micro-controllers to create more flexible and accurate nodes. With the greater availability of communication technologies combined with the work done to improve calibration of these sensor nodes, this paper shows the potential of this type of platform for wide-scale deployment.   

In this paper we will be presenting our cost-effective, agile, measurement platform. We will also present the results from co-location of this platform at the Cape Point climate measurement station in the Western Cape, South Africa. Finally, we will show the performance of different machine learning implementations at calibrating the data recorded by the measurement platform to that of the co-located reference sensor.

\section{Monitoring System and Co-location}
\subsection{System Design and Information}
In order to collect data, a measurement system was created. The platform is based on Espressif's ESP32 System on a Chip (SoC) \cite{esp32_Datasheet} due to its ease of access and cost-effective features. This allowed for creating a network connected, agile, measurement platform with the ability to connect to multiple sensors. The platform, being designed in South Africa, was designed with an on-board Uninterruptible Power Supply (UPS) so that the platform could continue to monitor in the event the mains power was disabled, as is the case during load-shedding. The platform is currently designed to break out to two hardware Universal Asynchronous Receiver/ Transmitter (UART) headers for two separate sensors. These two sensors were an MH-Z19C CO$_2$ Nondispersive Infrared (NDIR), measuring CO$_2$ parts per million (ppm), and a ZH03B laser-scattering Particulate Matter (PM) sensor, measuring micrograms per cubic metre ($\mu$g/m3), both made by Winsen Electronics Technology Co. Both of these sensors only report integer values and do not have a very high level of accuracy ,± 50ppm+5\% reading value and ±15 $\mu$g/m3 respectively, when compared to that of the reference equipment. The platform is not limited to these sensors as the board can support any hardware compatible sensor with the required software changes. This platform has been being operated since 08/12/2021 at the University of Cape Town and was deployed at the South African Weather Service Cape Point measurement station since 12/09/2022.

\subsection{Co-Location Site and Equipment}
The South African Weather Service Cape Point measurement station is located in the Cape Point nature reserve in Cape Town, South Africa. This measurement site works with the Global Monitoring Laboratory (GML) of the National Oceanic and Atmospheric Administration who specialize in research into: greenhouse gas and carbon cycle feedbacks, changes in clouds, aerosols, and surface radiation, and recovery of stratospheric ozone \cite{NOAA}. In this measurement site, the Picarro G2401 is used as the ground truth for the calibration of the test system and has an accuracy of 50, 20, or 10 parts per billion (ppb) depending on the chosen sample rate \cite{picarroG2401}. It is important to note that this measurement station is measuring atmospheric concentrations and is doing so by taking air from the top of a mast for their measurements. Our measurement platform was placed in an adjoining open-air room at the measurement facility and is subsequently more susceptible to local condition changes due to its location.

\section{Methodology}

In order to compare the results of the different calibration machine learning implementations, a set of metrics would need to be used. Theses metrics were chosen as they give insight into the different components of the data. We aim to quantify the performance with regard to the absolute values as well as the shape of the data. This will allow for better comparison and understanding of the results obtained during testing. The metrics we will use for comparison are Accuracy, Mean Absolute Error (MAE), R$^2$, as well as the Kullback-Leibler Divergence, seen in Equation \ref{eq:KLDiv}, and Jenson-Shannon Divergence, seen in Equation \ref{eq:JSDiv}. These metrics will allow for both spacial and probabilistic comparison to the reference data. The these can be described in their discrete forms, comparing two probabilistic distributions $P(x)$ and $Q(x)$, as seen in Nielsen\cite{nielsen2019jensen} as:

Kullback-Leibler Divergence
\begin{equation}
D_{K L}(P \| Q)=\sum_{x \in X} P(x) \ln \frac{P(x)}{Q(x)}
\label{eq:KLDiv}
\end{equation}

Entropy
\begin{equation}
H(X)=-\sum_{x \in \chi} P(x) \log _b P(x)
\label{eq:Entropy}
\end{equation}

Jenson-Shannon Divergence

\begin{equation}
\mathrm{JSD}(P \| Q)=\mathrm{H}\left(\frac{P+Q}{2}\right)-\frac{\mathrm{H}(P)+\mathrm{H}(Q)}{2}
\label{eq:JSDiv}
\end{equation}

\subsection{Description of the Data}
The collected data contains Date-Time, sensor system identification number, CO2 ppm and CO2 sensor temperature. The data received from the Cape Point does not contain the temperature of the sensor so it has been omitted from the work thus far. Due to the data from the Cape Point station being averaged for every minute, six values of our sensor system are used to predict one reading. This is due to the fact that our system is sampling every ten seconds. While there are months worth of readings collected, the testing was focused on the data from one day to determine the effectiveness of a relatively short period of calibration. This day contained 8525 measurements from our platform and 1432 readings from the instrument at Cape Point. This resulted in 1420 matched groups of 6 readings from our system per reading from Cape Point. The data is then split into a training set, accounting for 75\% of the data, and a test set, accounting for the remaining 25\%. This test/training data split is used across all of the machine learning methods tested.

\subsection{Random Forest Regression}
\begin{figure}[h]
    \centering
    \includegraphics[width=\columnwidth]{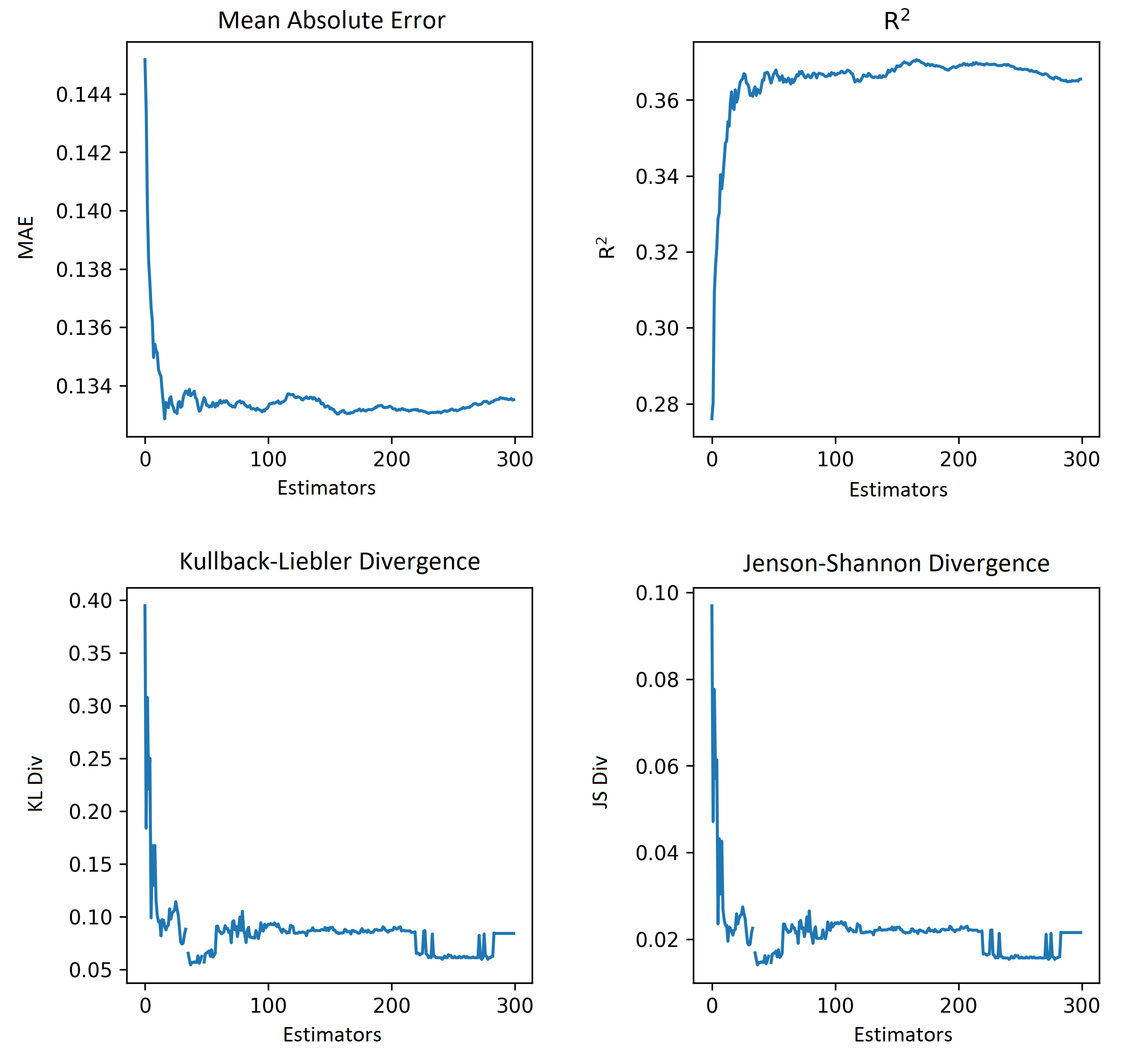}
    \caption{Figure showing random forest regression testing metrics per the number of estimators present in the model. The training shows heavily diminished returns above 15 estimators.}
    \label{fig:RFR-metrics-predicts}
\end{figure}

Random Forests for regression are formed by growing trees depending on a random vector that take on numerical values \cite{breiman2001random}.
The result returned by the Random Forest is none other than the average of the numerical result
returned by the different trees \cite{iannace2019wind}. To choose the best performance, a test was done to measure the performance relating to the number of estimators present in the model. The results can be seen in Fig.~\ref{fig:RFR-metrics-predicts}. The model chosen used 10 estimators, and was trained with bootstrapping.

\subsection{Artificial Neural Network}
 
As stated by Agatonovic-Kustrin et al \cite{agatonovic2000basic}, artificial neural networks (ANNs) are biologically inspired computer programs designed to simulate the way in which the human brain processes information. ANNs gather their knowledge by detecting the patterns and relationships in data and learn (or are trained) through experience, not from programming. This makes it an excellent candidate for calibrating data from sensors as it is able to learn how to correctly calibrate the input data through supervised learning. It also has the ability to handle non-linearity as it is not bound by linear modelling. For the model used in this paper, it was found that a three-layer sequential ANN would be used with 6, 150 and 1 nodes respectively. The tanh activation function was used.

\subsection{Support Vector Machine Regression}

Support Vector Regression (SVR) is a machine learning technique in which a model learns a variable's importance for characterizing the relationship between input and output \cite{zhang2020support}. One of the major tricks of Support Vector Machine (SVM) learning is the use of kernel functions to extend the class of decision functions to the non-linear case. This is done by mapping the data from the input space into a high-dimensional feature space as noted by R\"{u}ping\cite{Ruping2001kernels}. This made it an obvious candidate for testing and was implemented using a radial basis function (rbf) kernel as it performs well at a wide range of problems\cite{Ruping2001kernels}.

\section{Results}
The results of the experimentation done are compared in Table \ref{tab:ML-Results}. It indicates that the RFR implementation was able to perform the best overall. The predicted data is shown against the Cape Point data for each of the different implementations and the raw data collected from the cost-effective sensor platform.

\begin{table}[htbp]
\caption{Table showing the performance metrics of the different machine learning approaches used to calibrate the readings of the Cape Point co-located sensor node.}
\large
\resizebox{\columnwidth}{!}{%
\begin{tabular}{|c|c|c|c|c|c|}
\hline
\textbf{Method} & \textbf{Accuracy (\%)} & \textbf{MAE (ppm)} & \textbf{R$^2$} & \textbf{KL Divergence}  & \textbf{JS Divergence}  \\ \hline
Raw    & 94.82         & 21.51      & -7981                & 1.016  &  0.21 \\
RFR    & 99.97         & 0.14      & 0.34                 & 0.12  & 0.027 \\
ANN    & 99.96         & 0.18      & -0.0016              & -1.78 & 2.37  \\
SVR    & 99.97         & 0.13      & 0.36                & -1.78  & 2.37  \\ \hline
\end{tabular}%
}

\label{tab:ML-Results}
\end{table}

\subsection{Raw Data}
\begin{figure}[h]
    \centering
    \includegraphics[width=\columnwidth]{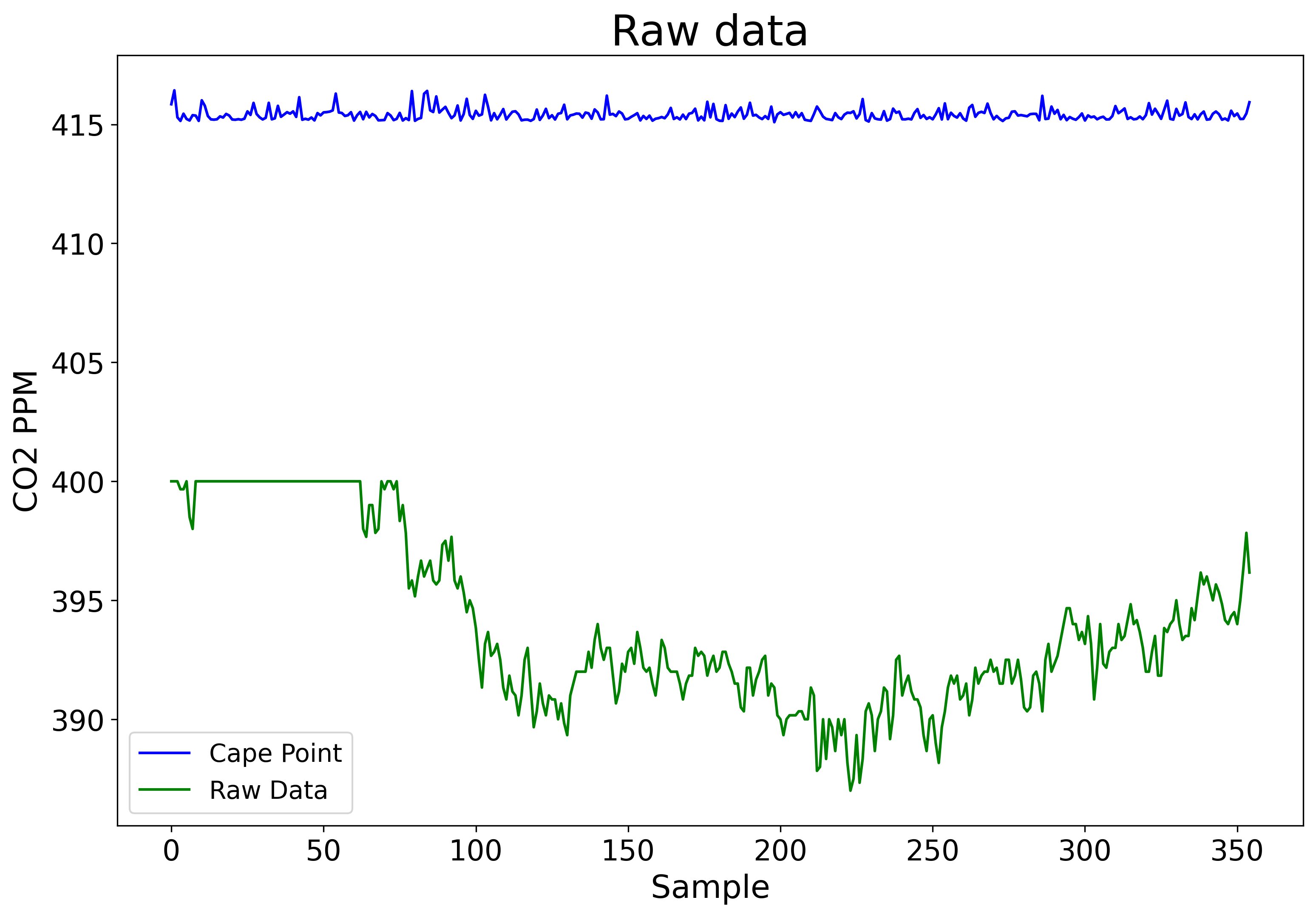}
    \caption{Figure showing Cape Point raw test data from our cost-effective platform.}
    \label{fig:RawTest}
\end{figure}

The raw data from the cost-effective platform placed at Cape Point was able to predict the data with an accuracy of 94.82\%. The sensor achieved a MAE of 21.51 ppm and an R$^2$ value of -7981, which is caused by the relatively high MAE. This was caused by the factory calibration of the sensor. The sensor showed a Kullback-Liebler divergence of 1.016 and a Jenson-Shannon divergence of 0.21. The recorded raw data can be seen in Fig.~\ref{fig:RawTest}.

\subsection{Random Forest Regression}
\begin{figure}[h]
    \centering
    \includegraphics[width=\columnwidth]{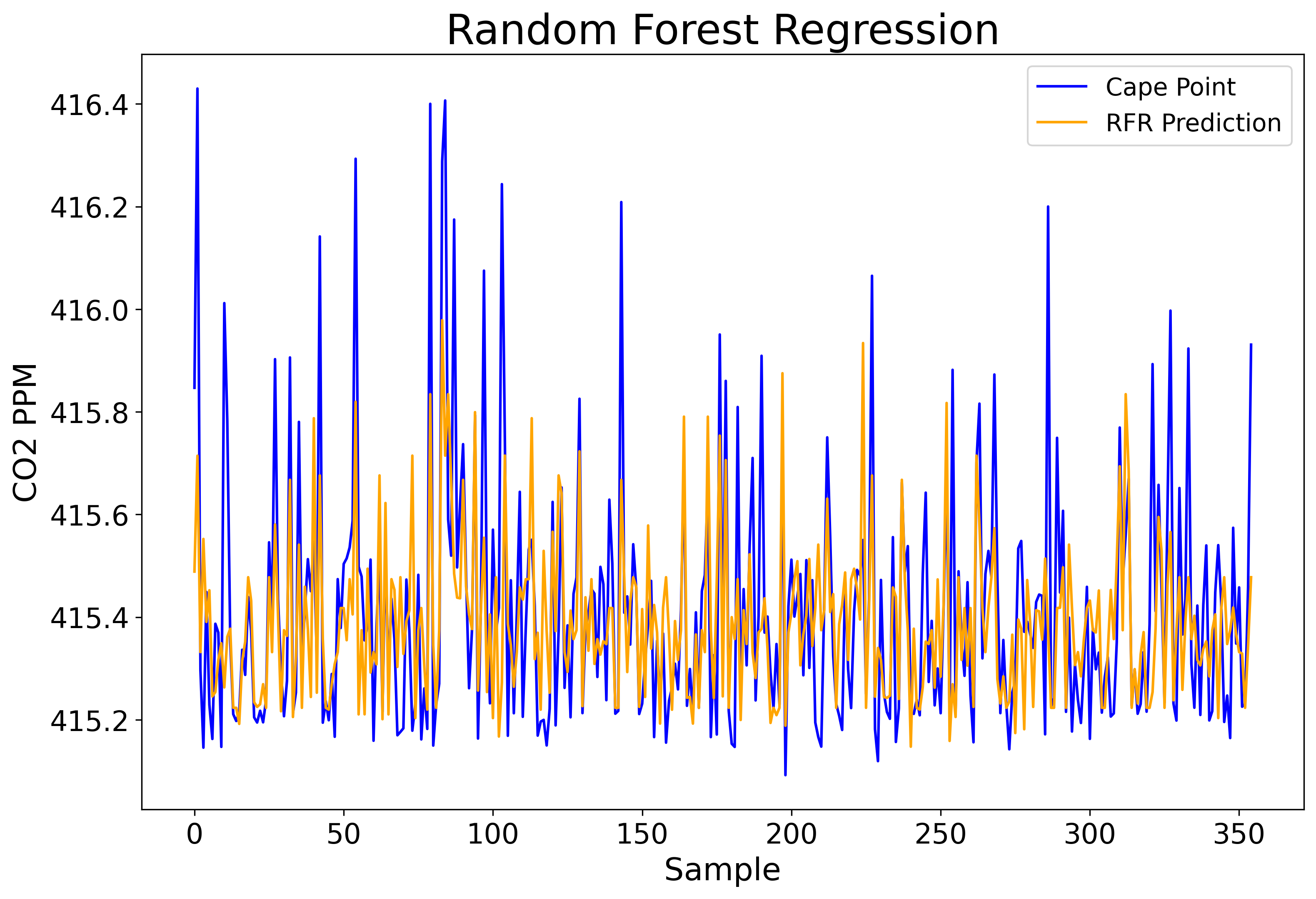}
    \caption{Figure showing RFR prediction of unseen Cape Point raw test data.}
    \label{fig:RFR-predicts}
\end{figure}

The Random Forest model was able to predict the data with an accuracy of 99.97\%. The model achieved a MAE of 0.14 ppm and an R$^2$ value of 0.34. The model was also able to achieve a Kullback-Liebler divergence of 0.12 and a Jenson-Shannon divergence of 0.027. This indicates that while the predictions share a high degree of similarity in their probability distributions, they do not have a high coefficient of determination. A low R$^2$ score implies that they do not have a strong correlation of best fit. The predictions made by this model can be seen in Fig.~\ref{fig:RFR-predicts}.

\subsection{Artificial Neural Network}
\begin{figure}[h]
    \centering
    \includegraphics[width=\columnwidth]{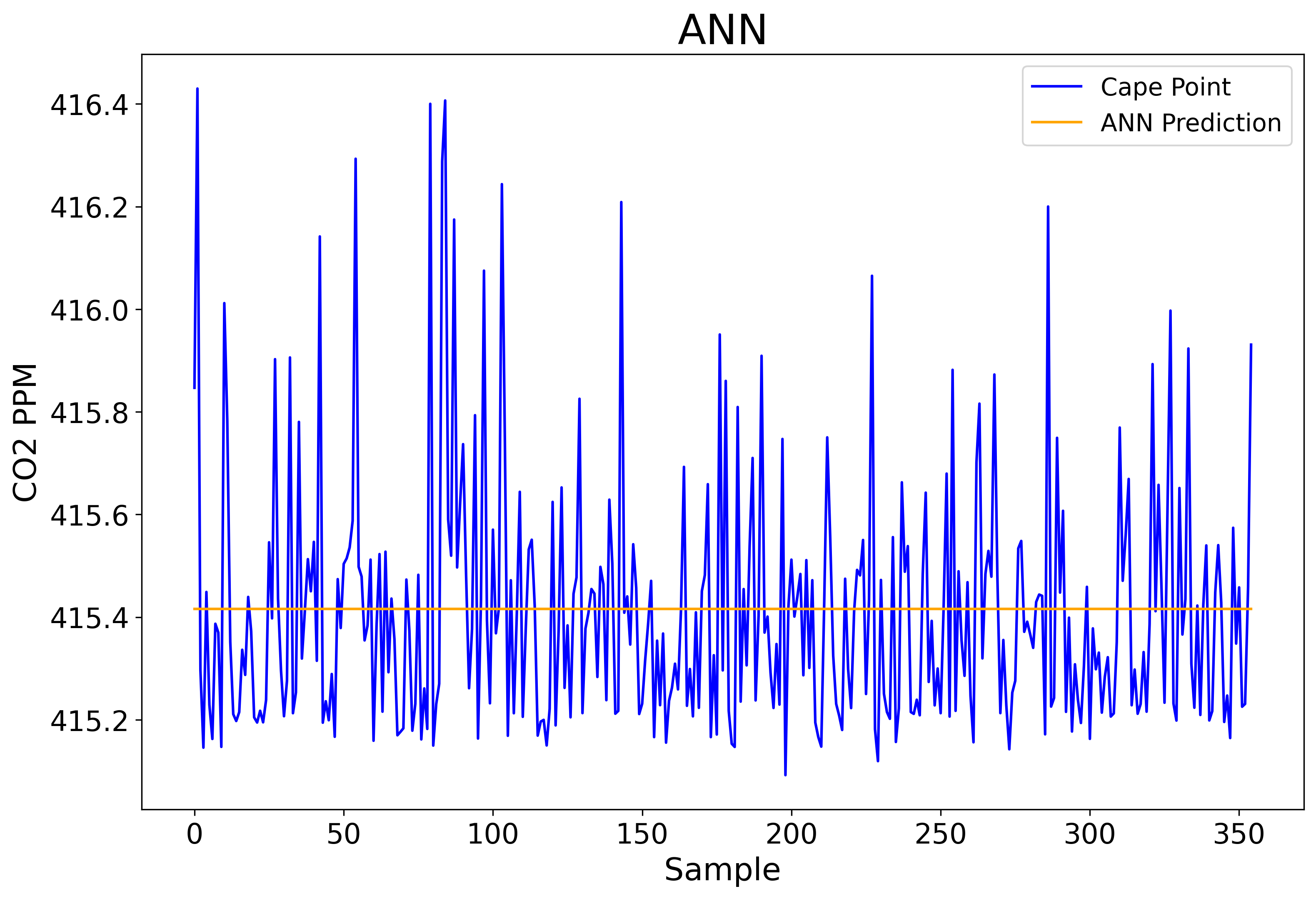}
    \caption{Figure showing ANN prediction of unseen Cape Point raw test data.}
    \label{fig:ANN-predicts}
\end{figure}

The ANN chosen was able to produce an accuracy of 99.96\% and a MSE of 0.18 ppm. It is interesting to note that the ANN produced this straight line for prediction. It scores terribly in the R$^2$ metric as well as in all the probability distribution metrics. The predictions made by this model can be seen in Fig.~\ref{fig:ANN-predicts}.

\subsection{Support Vector Machine Regression}
\begin{figure}[h]
    \centering
    \includegraphics[width=\columnwidth]{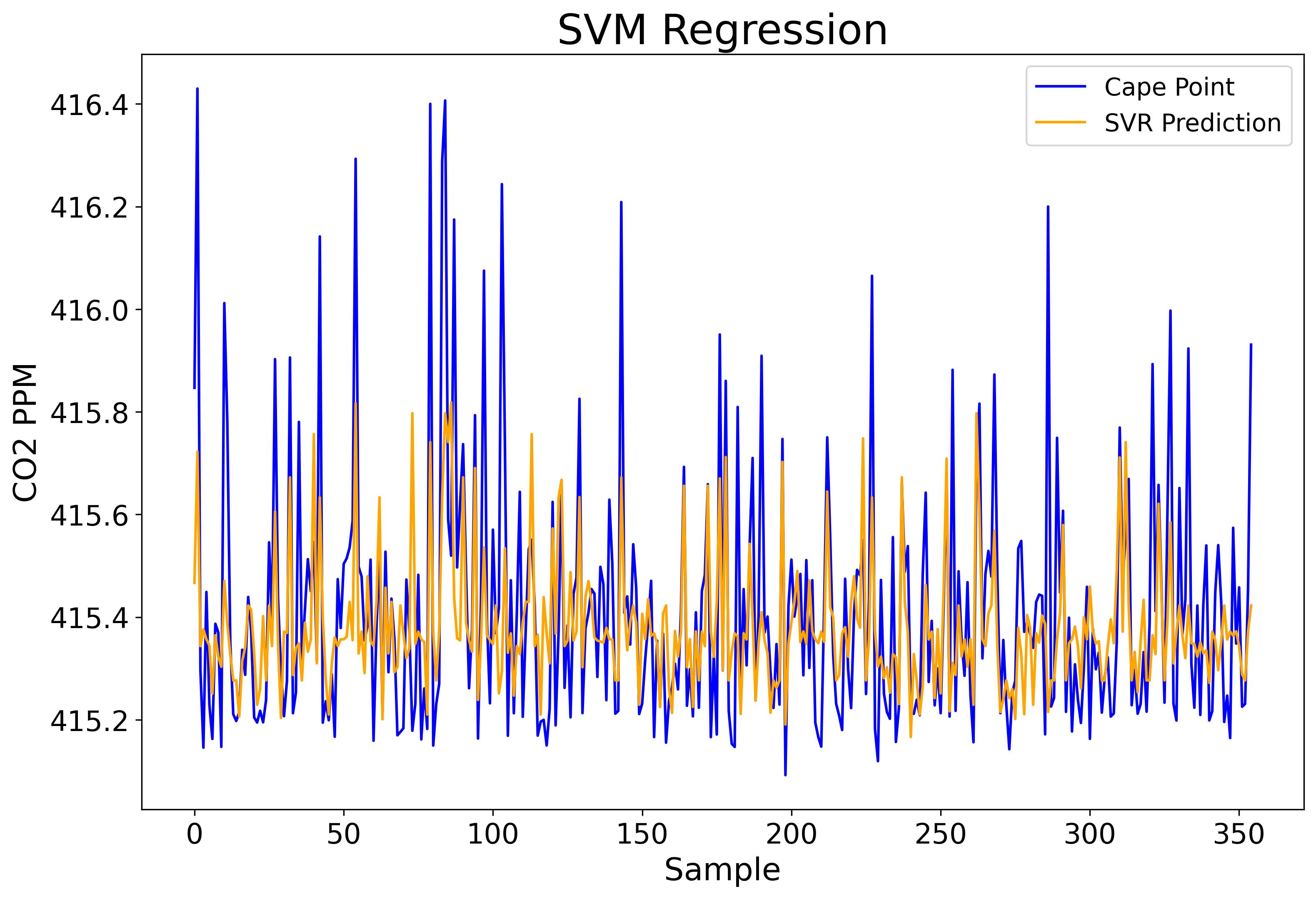}
    \caption{Figure showing SVR prediction of unseen Cape Point raw test data.}
    \label{fig:SVR-predicts}
\end{figure}

The SVM was able to achieve an accuracy of 99.97\% and a MAE of 0.13 ppm. It does not score well in the R$^2$ metric as well as in all the probability distribution metrics. The predictions made by this model can be seen in Fig.~\ref{fig:SVR-predicts}.

\section{Discussion}
While all the models tested have shown they have the ability to reduce the MAE of the data significantly, it is clear that only one implementation was able to closely replicate the probability distribution of the Cape Point data. This is shown clearly in Table \ref{tab:ML-Results}, indicating that all the machine learning methods have been able to greatly decrease the MAE over that of the raw data. However, only the RFR implementation was able to decrease the divergence between the probability distributions. This shows the promise of the machine learning approach to calibration of sensor data.

What is particularly noteworthy about these results is that the Cape Point data is consistent in the parts per million range. This is due to the location and purpose of this measurement facility. As the facility is primarily focused on atmospheric measurements, there is very little disturbance from the local environment. In order to find out the true performance of a system like this in an urban environment, it may be beneficial to find another measurement facility in an area with higher reading variability. This may lead to a better understanding of the actual performance of these machine learning methods.

\section{Conclusion}
In this paper we have shown the performance of the original readings from our cost-effective, agile, sensor platform. We have shown the performance of three different machine learning implementations, Random Forest Regression, Artificial Neural Network and Support Vector Machine Regression. It was found that the RFR was the best performing as it reduced the MAE to 0.14 while also achieving the best probability distribution metrics in our testing.

It was remarked upon that the reference sensor and measurement environment did not have a high level of variability and that it would be beneficial to test in an environment where variability is higher. 

From our testing it was clear that these machine learning implementations could improve the accuracy of our cost-effective platform and therefore, possibly, increase the time between manual calibration events in a sensor network.

\section*{Acknowledgment}
We would like to thank Sentech Soc Ltd, Weather South Africa and their team at Cape Point for their assistance with our research.

\bibliographystyle{IEEEtran}
\bibliography{Bib}

\end{document}